\documentclass[
 reprint,
 superscriptaddress,
frontmatterverbose, 
 amsmath,amssymb,
floatfix,
]{revtex4-2}
\usepackage[T1]{fontenc}
\usepackage[english]{babel}
\usepackage[makeindex]{splitidx}
\usepackage{xpatch}
\usepackage{graphicx}% Include figure files
\usepackage{dcolumn}% Align table columns on decimal point
\usepackage{bm}% bold math
\usepackage{color}

\usepackage{hyperref}% add hypertext capabilities
\begin{document}

%\preprint{APS/123-QED}

\title{First direct measurement of $^{22}{\rm Mg}(\alpha,p)^{25}{\rm Al}$ and implications for X-ray burst model-observation comparisons}% Force %line breaks with \\
%\thanks{A footnote to the article title}%
\author{J. S. Randhawa}\email{Corresponding author. randhawa@nscl.msu.edu}
%\email{randhawa@nscl.msu.edu}
\affiliation{National Superconducting Cyclotron Laboratory, Michigan State University, East Lansing, MI 48824, USA}
\affiliation{Joint Institute for Nuclear Astrophysics- Center for the Evolution of the Elements, Michigan State University,East Lansing,MI48824,USA}
%\email{randhawa@nscl.msu.edu}
\author{Y. Ayyad}\email{Corresponding author. ayyadlim@nscl.msu.edu}
%\email{ayyadlim@frib.msu.edu}
\affiliation{National Superconducting Cyclotron Laboratory, Michigan State University, East Lansing, MI 48824, USA}
%\affiliation{FRIB}
\author{W. Mittig}
\affiliation{National Superconducting Cyclotron Laboratory, Michigan State University, East Lansing, MI 48824, USA}
\affiliation{Department of Physics and Astronomy, Michigan State University, East Lansing, Michigan 48824-1321, USA}

\author{Z. Meisel}
\affiliation{Institute of Nuclear and Particle Physics, Department of Physics \& Astronomy, Ohio University, Athens, OH 45701, USA}
%\affiliation{INPP,OHio}
 \author{T. Ahn}
 \affiliation{Department of Physics, University of Notre Dame, Notre Dame, Indiana 46556-5670, USA}

 \author{S. Aguilar}
\affiliation{Department of Physics, University of Notre Dame, Notre Dame, Indiana 46556-5670, USA}
 \author{H. Alvarez-Pol}
 \affiliation{IGFAE, Universidade de Santiago de Compostela, E-15782, Santiago de Compostela, Spain}

 \author{D. W. Bardayan}
 \affiliation{Department of Physics, University of Notre Dame, Notre Dame, Indiana 46556-5670, USA}
 \author{D. Bazin}
 \affiliation{National Superconducting Cyclotron Laboratory, Michigan State University, East Lansing, MI 48824, USA}
\author{ S. Beceiro-Novo}
\affiliation{Department of Physics and Astronomy, Michigan State University, East Lansing, Michigan 48824-1321, USA}
\author{D. Blankstein}
  \affiliation{Department of Physics, University of Notre Dame, Notre Dame, Indiana 46556-5670, USA}
 \author{L. Carpenter}
 \affiliation{National Superconducting Cyclotron Laboratory, Michigan State University, East Lansing, MI 48824, USA}
 \author{ M. Cortesi}
 \affiliation{National Superconducting Cyclotron Laboratory, Michigan State University, East Lansing, MI 48824, USA}
  \author{D. Cortina-Gil}
  \affiliation{IGFAE, Universidade de Santiago de Compostela, E-15782, Santiago de Compostela, Spain}
  
  \author{P. Gastis}
  \affiliation{Central Michigan University, Mount Pleasant, Michigan 48859, USA}
  \affiliation{Joint Institute for Nuclear Astrophysics- Center for the Evolution of the Elements, Michigan State University,East Lansing,MI48824,USA}

 \author{M. Hall}
 \affiliation{Department of Physics, University of Notre Dame, Notre Dame, Indiana 46556-5670, USA}
\author{S. Henderson}
\affiliation{Department of Physics, University of Notre Dame, Notre Dame, Indiana 46556-5670, USA}

\author{J. J. Kolata}
\affiliation{Department of Physics, University of Notre Dame, Notre Dame, Indiana 46556-5670, USA}
%\author{W. Lynch}
%\affiliation{National Superconducting Cyclotron Laboratory, Michigan State University, East Lansing, MI 48824, USA}
%\affiliation{MSU}
\author{T. Mijatovic}
\affiliation{National Superconducting Cyclotron Laboratory, Michigan State University, East Lansing, MI 48824, USA}
\affiliation{Ru\dj{}er Bo\v{s}kovi\'{c} Institute, HR-10002 Zagreb, Croatia}
\author{F. Ndayisabye}
\affiliation{National Superconducting Cyclotron Laboratory, Michigan State University, East Lansing, MI 48824, USA}
\author{P. O'Malley}
\affiliation{Department of Physics, University of Notre Dame, Notre Dame, Indiana 46556-5670, USA}
\author{J. Pereira}
\affiliation{National Superconducting Cyclotron Laboratory, Michigan State University, East Lansing, MI 48824, USA}
\author{A. Pierre}
\affiliation{National Superconducting Cyclotron Laboratory, Michigan State University, East Lansing, MI 48824, USA}
\author{H. Robert}
\affiliation{National Superconducting Cyclotron Laboratory, Michigan State University, East Lansing, MI 48824, USA}
\author{C. Santamaria}
\affiliation{National Superconducting Cyclotron Laboratory, Michigan State University, East Lansing, MI 48824, USA}
\affiliation{Nuclear Science Division, Lawrence Berkeley National Laboratory, Berkeley, California 94720, USA}
\author{H. Schatz}
\affiliation{National Superconducting Cyclotron Laboratory,
Michigan State University, East Lansing, MI 48824, USA}
\affiliation{Joint Institute for Nuclear Astrophysics- Center for the Evolution of the Elements, Michigan State University,East Lansing,MI48824,USA}
\affiliation{Department of Physics and Astronomy, Michigan State University, East Lansing, Michigan 48824-1321, USA}

\author{J. Smith}
\affiliation{National Superconducting Cyclotron Laboratory, Michigan State University, East Lansing, MI 48824, USA}
%\affiliation{NSCL}
\author{N. Watwood}
\affiliation{National Superconducting Cyclotron Laboratory, Michigan State University, East Lansing, MI 48824, USA}
\author{J. C. Zamora}
\affiliation{National Superconducting Cyclotron Laboratory, Michigan State University, East Lansing, MI 48824, USA}
\affiliation{Instituto de Fisica, Universidade de Sao Paulo, 05508-090 Sao Paulo, Brazil}
%\affiliation{Sau Paulo}

 %\altaffiliation[Also at ]{Physics Department, XYZ University.}%Lines break automatically or can be forced with \\

\date{ January 2020}% It is always \today, today,
             %  but any date may be explicitly specified

\begin{abstract}
Type-I X-ray burst (XRB) light curves are sensitive to the model's nuclear input, consequently affects the model-observation comparisons. $^{22}{\rm Mg}(\alpha,p)^{25}{\rm Al}$ is among the most important reactions that directly impact the XRB light curve. We report the first direct measurement of $^{22}{\rm Mg}(\alpha,p)^{25}{\rm Al}$ using the Active Target Time Projection Chamber. XRB light curve model-observation comparisons for the source $\tt{GS 1826-24}$ using new reaction rate imply a less-compact neutron star than previously inferred. Additionally, our result removes an important uncertainty in XRB model calculations.

%\begin{description}
%\item[Usage]

%\item[Structure]

%\end{description}
\end{abstract}

%\keywords{Suggested keywords}%Use showkeys class option if keyword
 \maketitle                             %display desired

%\tableofcontents

\section*{\label{sec:level1} }
Type-I X-ray bursts (XRBs) are the thermonuclear explosions on the surface of accreting neutron stars powered by the nuclear burning \cite{Lewin93,sch06,Jose16}. In recent years,  advances in XRB observations and modeling have opened a unique window to constrain the mass-radius relation and other underlying physics through comparisons between observations and models \cite{Meisel_2018,Meis18}. As XRB light curves are powered by nuclear reactions, XRB models are sensitive to the various nuclear physics inputs (e.g. nuclear reaction rates) \cite{Cybu10,Cyburt2016}. Models with reliable nuclear physics data are needed to validate the assumptions of the astrophysical models through model-observation comparisons \cite{Meisel_2019}. Accurate nuclear physics input plays an equally important role in predicting the burst ashes, which alter the composition of the crust of the neutron star, which in mass-accreting systems is made in part or entirely out of XRB ashes. Various sensitivity studies over the years have shown that the $^{22}{\rm Mg}(\alpha,p)^{25}{\rm Al}$ reaction rate is among the most significant reactions that directly impact the light curves and burst ashes \cite{Parikh_2008,Cybu10,Cyburt2016}. Recently in a study to assess the impact of uncertainties in nuclear inputs on the extraction of the neutron star mass-radius relation, the $^{22}{\rm Mg}(\alpha,p)^{25}{\rm Al}$ reaction rate was shown to have a significant effect even when decreased by a factor of 10 \cite{Meisel_2019}. \\

In XRBs at  temperatures 0.5-0.6 GK, breakout from the CNO cycle via $^{15}{\rm O}(\alpha,\gamma)^{19}{\rm Ne}$ and $^{18}{\rm Ne}(\alpha,p)^{21}{\rm Ne}$ becomes efficient. These breakout reactions open the door for the $\alpha p$-process, and the reaction flow reaches $^{22}$Mg. At this branching point, $^{22}{\rm Mg}(\alpha,p)^{25}{\rm Al}$ competes with the rather slow $\beta^{+}$-decay and with $^{22}{\rm Mg}(p,\gamma)^{23}{\rm Al}$ \cite{Meisel_2018}. The current experimental constraint on this reaction rate comes from an indirect measurement where resonant states in $^{26}$Si were explored through the $^{28}$Si(p,t)$^{26}$Si reaction by \citet{Matic_2011}. This experimentally constrained reaction rate is more than a factor of 100 below Hauser-Feshbach (HF) predictions in the relevant XRB temperature range above 0.7 GK. The large deviation from the HF based model calculations was attributed to the lack of resonance data above 10-MeV excitation energy in $^{26}$Si \cite{Matic_2011}; therefore their rate was considered to be a lower limit, and the HF based rate an upper limit. These two rates lead to significantly different results when used in XRB model calculations resulting in a significant uncertainty of model-observation comparisons. Since XRB ashes ultimately set the composition of the accreted neutron star crust, the discrepant ash results may also impact model-observation comparisons for neutron star crust cooling \cite{Meisel_2017,Meisel_2018,Lau18}. Therefore, it is important to directly measure this reaction to reduce this very large uncertainty and constrain the XRB model calculations. We report the first direct measurement of the $^{22}{\rm Mg}(\alpha,p)^{25}{\rm Al}$ reaction using the Active-Target Time Projection Chamber (AT-TPC).\\ 

\begin{figure}
    \centering
    \includegraphics[scale=0.6]{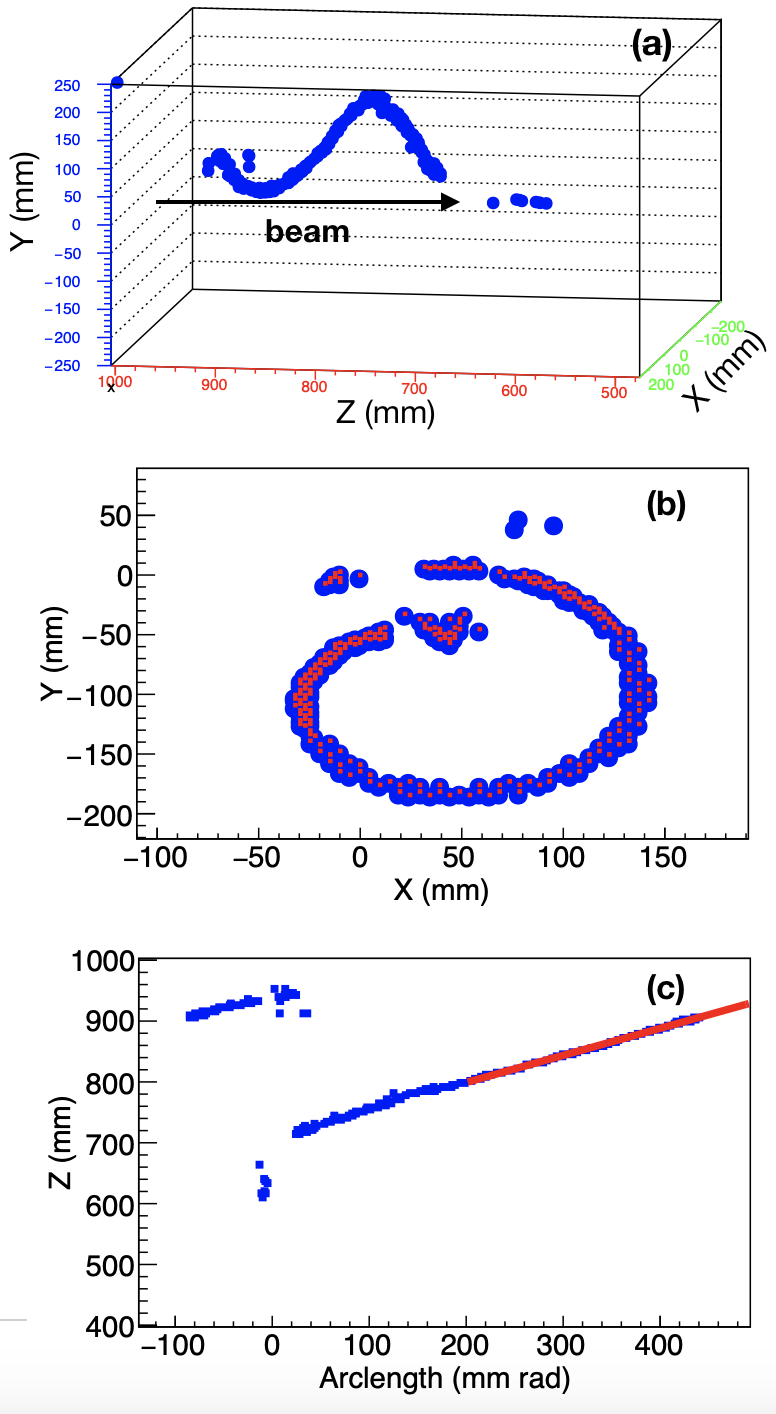}
    \caption{The top panel shows a three-dimensional view of an example proton track and the arrow shows the beam direction. The middle panel shows the projection of a proton track on the pad plane where blue dots are data points and red indicates the data points chosen for RANSAC analysis. The lower panel shows the arc length of each hit pattern point as a function of the z-coordinate. The red line is the least-squares fit performed to extract the scattering angle from the slope.}
    \label{fig1}
\end{figure}
 The $^{22}{\rm Mg}(\alpha,p)^{25}{\rm Al}$ measurement was carried out at the National Superconducting Cyclotron Laboratory (NSCL). $^{22}$Mg was produced from the fragmentation of a $^{24}$Mg primary beam accelerated by the coupled cyclotrons and selected by the A1900 fragment separator \cite{Morri2003}. The $^{22}$Mg fragments were stopped in a linear gas cell and transported to an electron beam ion source (EBIS), where their charge state increased to 12$^{+}$. Finally, the ions were injected into the ReA3 re-accelerator, accelerated to $\sim$5 MeV/u with an average beam intensity of $\sim$900 pps. The re-accelerated beam was transported through a thin ionization chamber filled with isobutane at 10 torr. The ionization chamber records the beam intensity and identifies beam contaminants \cite{Bradt2017}. Before the beam entered the active volume of the AT-TPC, it went through a 3.6 $\mu m$ thick aluminized para-aramid entrance window of 1 cm diameter. The AT-TPC active volume is a cylinder of length 1 m and of a radius 29.2 cm, placed in a uniform 1.9 T magnetic field generated by a solenoidal magnet. The AT-TPC was filled with 600 torr He:CO$_{2}$(95\%:5\%) to stop the beam in the middle of the AT-TPC. The sensor plane consisted of a mosaic of 10240 equilateral triangle pads and provides x and y information of the tracks. Drift time provides the  longitudinal component of the track. The ion chamber signal was used to retain the arrival of beam particles through the window as a time reference, which is pivotal to determine the reaction vertex position along the beam axis. Details about the electronics and trigger setup can be found in reference \cite{Bradt2017}. Among the dominating channels open at this energy are ($\alpha$,$\alpha$) and ($\alpha,$p). From kinematics, the only particles which were also back-angle emitted, i.e. $\theta_{\rm lab} >$ 90 degrees, are protons. Therefore protons were identified by selecting the back-scattered single tracks. Another possible source of back-scattered protons are fusion evaporation reactions on carbon and oxygen. Any background contribution from reactions on carbon in this angular domain were estimated (using PACE4 \cite{Tarasov2008}) to be less than 0.1\%. The back-scattered proton tracks were analyzed using the Random Sample consensus (RANSAC) method \cite{Ayyad2018,Ayyad_2018}.  Figure~\ref{fig1} shows an example proton track, where the top panel shows a back-scattered proton in 3D and the lower panel shows a 2D projection of the same track on the pad plane as well as RANSAC analysis. If the scattering angle was above 90 degrees, the track was selected for further analysis to obtain the reaction vertex. RANSAC uses a mathematical model (circle) to describe a collection of points. This allows for a determination of the radius of curvature since the first part of the spiral can be approximated by a circle. Once the radius of curvature is determined, the scattering angle can be inferred by parameterizing the position along the $z$-axis as a function of the arc length as shown in Figure~(\ref{fig1}c).

\begin{figure}
    \centering
    \includegraphics[scale=0.62]{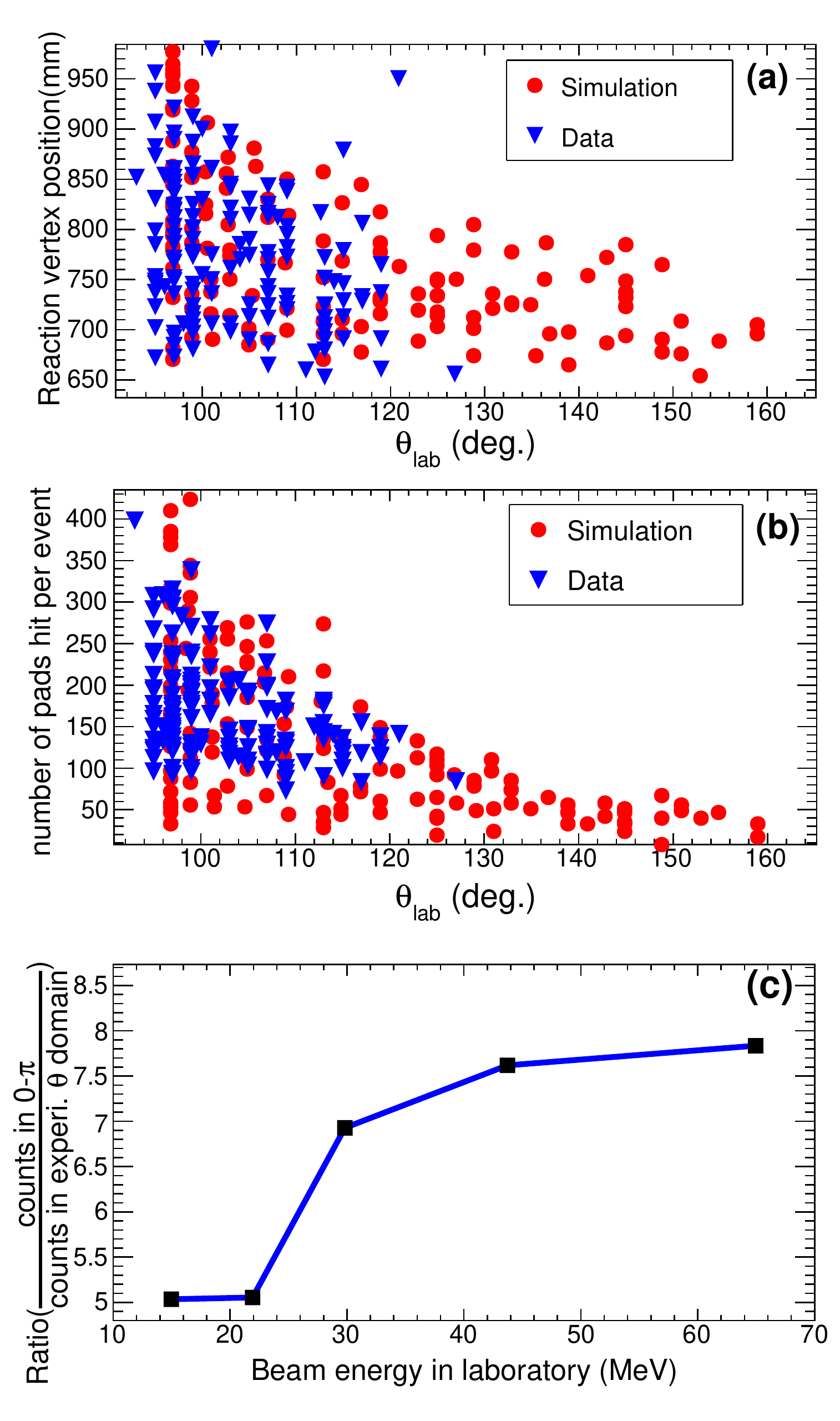}
    \caption{Panel (a) shows the reaction vertex as a function of laboratory angle for experimental data (blue inverted triangle) overlaid on simulation data (red filled circles). The middle panel shows the number of pads hit per event as a function of the laboratory angle. Panel (c) shows the ratio of calculated (using PACE4) total counts over the counts in the experimental angular domain as a function of the beam energy.}
    \label{figure2}
\end{figure}

Tracks for the ($\alpha$,p) reaction channel were simulated using GEANT4 \cite{Geant4,geant41} and were digitized to include the detailed detector and electronics effects. More about the AT-TPC simulation package and digitization can be found in \citet{Ayyad2018}. Simulated tracks were also analyzed as described above. The top panel of Figure~\ref{figure2} shows the event-by-event reaction vertex as a function of laboratory angle for the experimental and simulated data. It is evident from the plot that the experimentally accessible angular domain is limited from 90$^{\circ}$ to 130$^{\circ}$ in the laboratory frame. As we are analyzing only the backscattered protons, the higher angles are accessible only when the reaction vertex is at some distance from the entrance window. The detected angular range during the experiment depends on the geometrical acceptance of the detector and the threshold of the multiplicity trigger. In the current work, the multiplicity is defined as the number of pads fired in a given time window which depends on the angle of the reaction product. Figure ~\ref{figure2}b shows the multiplicity or pads fired per event as a function of laboratory angle. In simulations, the multiplicity threshold was set to zero and angles up to 160$^{\circ}$ can be seen. The experimental data show a sharp cut-off at $\sim$80 hits per event. Higher laboratory angles were cut off due to this multiplicity threshold in the current experiment. To obtain the angle-integrated cross-sections, the proton distribution in the laboratory frame was calculated at different beam energies using PACE4 \cite{Tarasov2008}. The ratio of counts from $ 0 - 180$ degrees to counts in the angular region covered in this study are shown in Figure~\ref{figure2}c. This energy-dependent ratio was used to obtain the angle integrated cross-section. \\
\begin{figure}
    \centering
    \includegraphics[scale=0.60]{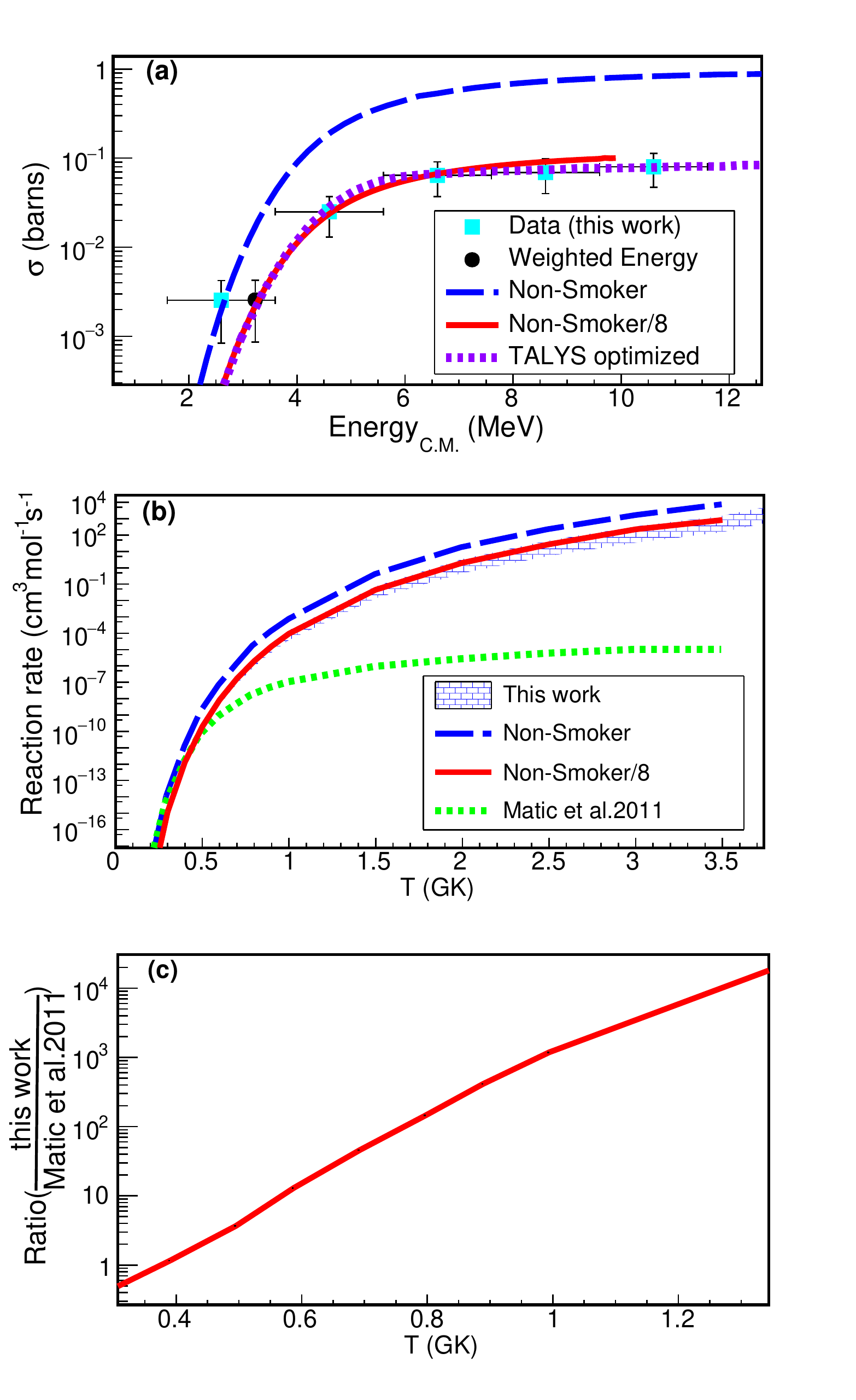}
    \caption{Panel(a) shows the experimental cross sections obtained in the present work over a range of center-of-mass energies covered (cyan color). For the lowest energy point, the black point represents the cross section weighted energy. The middle panel shows the reaction rate comparison of the current work to different model predictions and to the previous measurement by \citet{Matic_2011}. The lowest panel shows the ratio of present reaction rate to \citet{Matic_2011}.}
    \label{figure3}
\end{figure}
 The excitation function obtained is shown in the Figure 3a in comparison to HF calculations. Vertical error bars include contributions from statistical and systematic uncertainty. Systematic error bars include a 5\%  uncertainty in the number of target atoms based on uncertainties in the energy loss tables, 5\% in the incident beam counts, and 35\% error when accounting for counts outside the angular domain covered in this study. Our estimation of 35\% uncertainty is based on any uncertainty originating from the model prediction of the proton angular distribution. Horizontal error bars reflect the bin size in center-of-mass energies. The cross sections obtained in the current work are a factor of 8 to 10 lower than the HF calculation results. The lowest center-of-mass energy for our measurement is located near the upper-end of the astrophysical Gamow window for a 2 GK  temperature~\cite{Raus10}. Extrapolation to lower center-of-mass energies is required to obtain the cross section and reaction rate in the relevant temperature range. For this, a combination of input parameters for TALYS was identified, mainly double-folding alpha potential and the level density of the compound nucleus, to best reproduces the experimental data within error bars and is shown in Figure ~\ref{figure3}a. The optimized TALYS reaction rate is shown in Figure ~\ref{figure3}b and the shaded area shows the estimated reaction rate uncertainty. The energy dependence of the experimental data is consistent with HF model calculations. The dramatic fall-off in the reaction rate above 0.4 GK as seen in the previous measurement was not observed in the current work. With the current measurement the reaction rate in the critical temperature range around 0.7-1 GK is determined experimentally for the first time, and its uncertainty is dramatically reduced.

\begin{figure}
    \centering
    \includegraphics[scale=0.51]{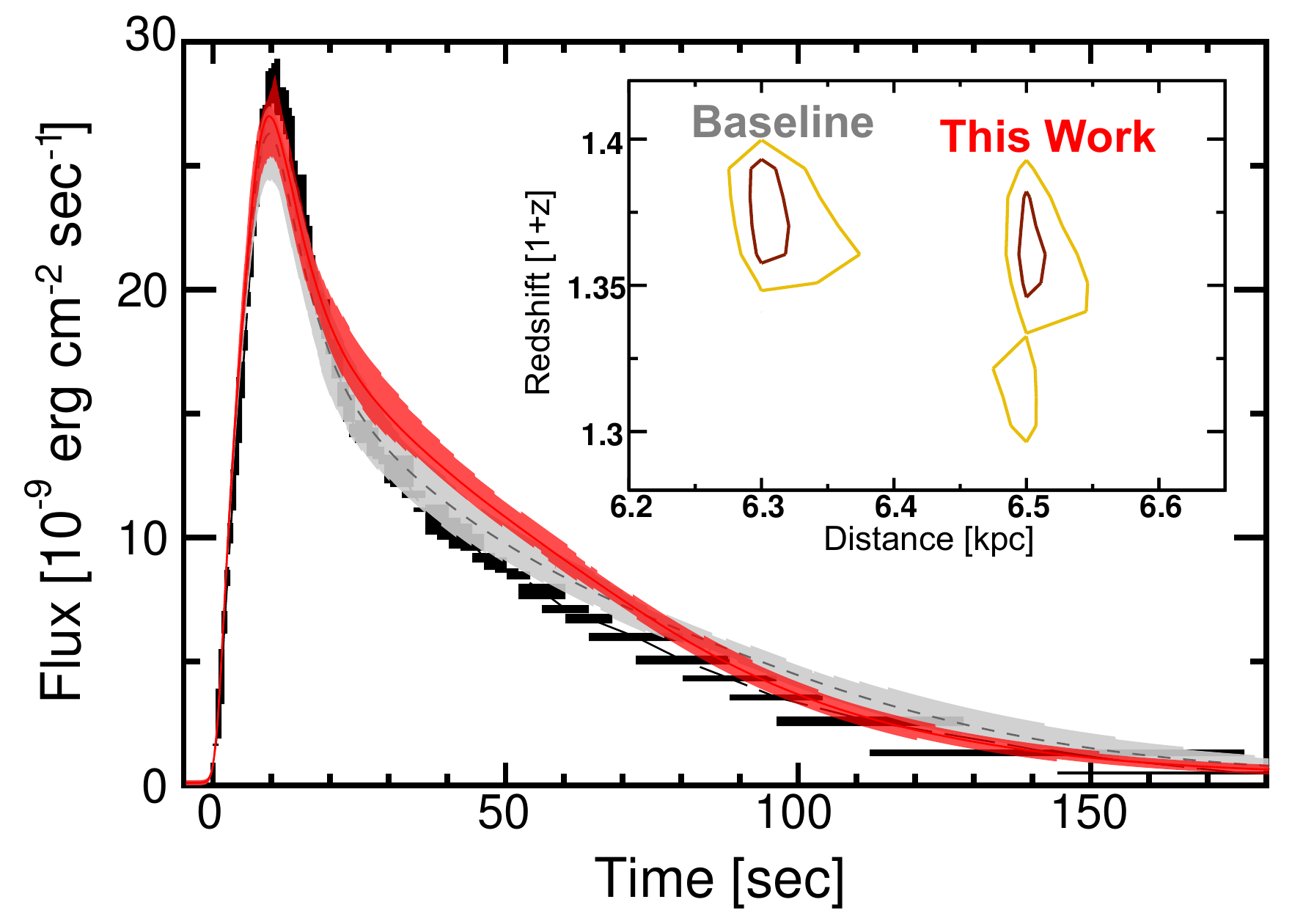}
    \caption{XRB light curve comparison for GS 1826-24 observations (black boxes), {\tt MESA} calculations using the {\tt NON-SMOKER} rate for $^{22}{\rm Mg}(\alpha,p)$ (gray band), and the {\tt NON-SMOKER} rate reduced by $/8$ (red band). The inset shows the corresponding 68\% (red) and 95\% (yellow) confidence intervals for the distance and redshift, which were determined as described in Reference~\citep{Meisel_2019}.}
    \label{fig:my_label}
\end{figure}

To assess the impact of our measurement on the XRB light curve, we performed multizone XRB calculations with the code {\tt MESA}, following those described in References~\cite{Meis18,Meisel_2019}. We employ the REACLIBv2.2 nuclear reaction rate library, where the {\tt NON-SMOKER} HF rate of Reference~\cite{Cybu10} is the default for $^{22}{\rm Mg}(\alpha,p)^{25}{\rm Al}$ . We used the astrophysical conditions that were found to best reproduce observables from the year 2000 bursting epoch of the source GS 1826-24~\cite{Gall17,Meis18}. Figure~\ref{fig:my_label} compares {\tt MESA} results to astronomical observations. We use a distance of 6.2~kpc and redshift ($1+z$) of 1.38, which provide the best-fit between the observed light curve and our baseline calculation, to mimic the X-ray detection solid angle and neutron star surface gravitational redshift that modify the observed light curve.

Figure \ref{fig:my_label} shows that the calculations from References~\cite{Meis18,Meisel_2019} (gray band) generally reproduce the observed light curve (black boxes), including the recurrence time between bursts (which is not shown). However, this agreement is substantially diminished when reducing the {\tt NON-SMOKER} rate for $^{22}{\rm Mg}(\alpha,p)$ by a factor of 8 (red band). In particular, the tail of the XRB light curve is substantially modified because a reduced $^{22}{\rm Mg}(\alpha,p)$ rate effectively enhances hydrogen burning early in the burst by making the path $^{22}{\rm Mg}(p,\gamma)^{23}{\rm Al}(p,\gamma)^{24}{\rm Si}$ \cite{Wolf19} more competitive. Therefore less hydrogen is available to be burned at later times following the light curve peak, resulting in a more rapid decline of the light curve tail. The implication of our result is that an alternative distance and surface gravitational redshift are needed in order to reproduce observed features of GS 1826-24, as shown in the Figure~\ref{fig:my_label} inset. This implies a less-compact neutron star in GS 1826-24 than determined by Reference~\cite{Meisel_2019}. Significantly, our result removes an important uncertainty in XRB model calculations that previously hindered extraction of neutron star compactness via XRB light curve model-observation comparisons.

%\color{black}
\acknowledgements{ Authors would like to thank the beam delivery group. This work was supported by the U.S.  National  Science  Foundation (NSF), USA under grants MRI-0923087 and PHY-140444, under Cooperative Agreement No. PHY-1565546, Grant No. PHY-1713857 and by  award numbers PHY-1430152(JINA Center for the Evolution of the Elements), PHY-1102511, and PHY-1913554. T.A. and D.W.B. acknowledges NSF grant no. PHY-1713857, J.J.K. acknowledges the NSF grant no. 14-01343. P.G. would like to acknowledge the support by the College of Science \& Engineering of CMU. Z.M. acknowledges the U.S. Department of Energy under grants DE-FG02-88er40387 and DESC0019042. J.C.Z. thanks the support by  Fundação  de  Amparo  a Pesquisa do Estado de São Paulo (FAPESP) under Grant No. 2018/04965-4. Authors are thankful to Domenico Santonocito (INFN) for providing the Cascade calculations.}

\bibliography{main}% Produces the bibliography via BibTeX.

%apsrev4-2.bst 2019-01-14 (MD) hand-edited version of apsrev4-1.bst
%Control: key (0)
%Control: author (8) initials jnrlst
%Control: editor formatted (1) identically to author
%Control: production of article title (0) allowed
%Control: page (0) single
%Control: year (1) truncated
%Control: production of eprint (0) enabled
\begin{thebibliography}{22}%
\makeatletter
\providecommand \@ifxundefined [1]{%
 \@ifx{#1\undefined}
}%
\providecommand \@ifnum [1]{%
 \ifnum #1\expandafter \@firstoftwo
 \else \expandafter \@secondoftwo
 \fi
}%
\providecommand \@ifx [1]{%
 \ifx #1\expandafter \@firstoftwo
 \else \expandafter \@secondoftwo
 \fi
}%
\providecommand \natexlab [1]{#1}%
\providecommand \enquote  [1]{``#1''}%
\providecommand \bibnamefont  [1]{#1}%
\providecommand \bibfnamefont [1]{#1}%
\providecommand \citenamefont [1]{#1}%
\providecommand \href@noop [0]{\@secondoftwo}%
\providecommand \href [0]{\begingroup \@sanitize@url \@href}%
\providecommand \@href[1]{\@@startlink{#1}\@@href}%
\providecommand \@@href[1]{\endgroup#1\@@endlink}%
\providecommand \@sanitize@url [0]{\catcode `\\12\catcode `\$12\catcode
  `\&12\catcode `\#12\catcode `\^12\catcode `\_12\catcode `\%12\relax}%
\providecommand \@@startlink[1]{}%
\providecommand \@@endlink[0]{}%
\providecommand \url  [0]{\begingroup\@sanitize@url \@url }%
\providecommand \@url [1]{\endgroup\@href {#1}{\urlprefix }}%
\providecommand \urlprefix  [0]{URL }%
\providecommand \Eprint [0]{\href }%
\providecommand \doibase [0]{https://doi.org/}%
\providecommand \selectlanguage [0]{\@gobble}%
\providecommand \bibinfo  [0]{\@secondoftwo}%
\providecommand \bibfield  [0]{\@secondoftwo}%
\providecommand \translation [1]{[#1]}%
\providecommand \BibitemOpen [0]{}%
\providecommand \bibitemStop [0]{}%
\providecommand \bibitemNoStop [0]{.\EOS\space}%
\providecommand \EOS [0]{\spacefactor3000\relax}%
\providecommand \BibitemShut  [1]{\csname bibitem#1\endcsname}%
\let\auto@bib@innerbib\@empty
%</preamble>
\bibitem [{\citenamefont {{Lewin}}\ \emph {et~al.}(1993)\citenamefont
  {{Lewin}}, \citenamefont {{van Paradijs}},\ and\ \citenamefont
  {{Taam}}}]{Lewin93}%
  \BibitemOpen
  \bibfield  {author} {\bibinfo {author} {\bibfnamefont {W.~H.~G.}\
  \bibnamefont {{Lewin}}}, \bibinfo {author} {\bibfnamefont {J.}~\bibnamefont
  {{van Paradijs}}},\ and\ \bibinfo {author} {\bibfnamefont {R.~E.}\
  \bibnamefont {{Taam}}},\ }\bibfield  {title} {\bibinfo {title} {{X-Ray
  Bursts}},\ }\href {https://doi.org/10.1007/BF00196124} {\bibfield  {journal}
  {\bibinfo  {journal} {Space Science Review}\ }\textbf {\bibinfo {volume}
  {62}},\ \bibinfo {pages} {223} (\bibinfo {year} {1993})}\BibitemShut
  {NoStop}%
\bibitem [{\citenamefont {Schatz}\ and\ \citenamefont {Rehm}(2006)}]{sch06}%
  \BibitemOpen
  \bibfield  {author} {\bibinfo {author} {\bibfnamefont {H.}~\bibnamefont
  {Schatz}}\ and\ \bibinfo {author} {\bibfnamefont {K.}~\bibnamefont {Rehm}},\
  }\bibfield  {title} {\bibinfo {title} {X-ray binaries},\ }\href
  {https://doi.org/https://doi.org/10.1016/j.nuclphysa.2005.05.200} {\bibfield
  {journal} {\bibinfo  {journal} {Nuclear Physics A}\ }\textbf {\bibinfo
  {volume} {777}},\ \bibinfo {pages} {601 } (\bibinfo {year}
  {2006})}\BibitemShut {NoStop}%
\bibitem [{\citenamefont {{Jose}}(2016)}]{Jose16}%
  \BibitemOpen
  \bibfield  {author} {\bibinfo {author} {\bibfnamefont {J.}~\bibnamefont
  {{Jose}}},\ }\href {https://doi.org/10.1201/b19165} {\emph {\bibinfo {title}
  {Stellar Explosions: Hydrodynamics and Nucleosynthesis by Jordi Jose}}}\
  (\bibinfo  {publisher} {CRC Press/Taylor and Francis},\ \bibinfo {year}
  {2016})\BibitemShut {NoStop}%
\bibitem [{\citenamefont {Meisel}\ \emph {et~al.}(2018)\citenamefont {Meisel},
  \citenamefont {Deibel}, \citenamefont {Keek}, \citenamefont {Shternin},\ and\
  \citenamefont {Elfritz}}]{Meisel_2018}%
  \BibitemOpen
  \bibfield  {author} {\bibinfo {author} {\bibfnamefont {Z.}~\bibnamefont
  {Meisel}}, \bibinfo {author} {\bibfnamefont {A.}~\bibnamefont {Deibel}},
  \bibinfo {author} {\bibfnamefont {L.}~\bibnamefont {Keek}}, \bibinfo {author}
  {\bibfnamefont {P.}~\bibnamefont {Shternin}},\ and\ \bibinfo {author}
  {\bibfnamefont {J.}~\bibnamefont {Elfritz}},\ }\href
  {https://doi.org/10.1088/1361-6471/aad171} {\bibfield  {journal} {\bibinfo
  {journal} {Journal of Physics G: Nuclear and Particle Physics}\ }\textbf
  {\bibinfo {volume} {45}},\ \bibinfo {pages} {093001} (\bibinfo {year}
  {2018})}\BibitemShut {NoStop}%
\bibitem [{\citenamefont {Meisel}(2018)}]{Meis18}%
  \BibitemOpen
  \bibfield  {author} {\bibinfo {author} {\bibfnamefont {Z.}~\bibnamefont
  {Meisel}},\ }\href {https://doi.org/10.3847/1538-4357/aac3d3} {\bibfield
  {journal} {\bibinfo  {journal} {The Astrophysical Journal}\ }\textbf
  {\bibinfo {volume} {860}},\ \bibinfo {pages} {147} (\bibinfo {year}
  {2018})}\BibitemShut {NoStop}%
\bibitem [{\citenamefont {Cyburt}\ \emph {et~al.}(2010)\citenamefont {Cyburt},
  \citenamefont {Amthor}, \citenamefont {Ferguson}, \citenamefont {Meisel},
  \citenamefont {Smith}, \citenamefont {Warren}, \citenamefont {Heger},
  \citenamefont {Hoffman}, \citenamefont {Rauscher}, \citenamefont {Sakharuk},
  \citenamefont {Schatz}, \citenamefont {Thielemann},\ and\ \citenamefont
  {Wiescher}}]{Cybu10}%
  \BibitemOpen
  \bibfield  {author} {\bibinfo {author} {\bibfnamefont {R.~H.}\ \bibnamefont
  {Cyburt}}, \bibinfo {author} {\bibfnamefont {A.~M.}\ \bibnamefont {Amthor}},
  \bibinfo {author} {\bibfnamefont {R.}~\bibnamefont {Ferguson}}, \bibinfo
  {author} {\bibfnamefont {Z.}~\bibnamefont {Meisel}}, \bibinfo {author}
  {\bibfnamefont {K.}~\bibnamefont {Smith}}, \bibinfo {author} {\bibfnamefont
  {S.}~\bibnamefont {Warren}}, \bibinfo {author} {\bibfnamefont
  {A.}~\bibnamefont {Heger}}, \bibinfo {author} {\bibfnamefont {R.~D.}\
  \bibnamefont {Hoffman}}, \bibinfo {author} {\bibfnamefont {T.}~\bibnamefont
  {Rauscher}}, \bibinfo {author} {\bibfnamefont {A.}~\bibnamefont {Sakharuk}},
  \bibinfo {author} {\bibfnamefont {H.}~\bibnamefont {Schatz}}, \bibinfo
  {author} {\bibfnamefont {F.~K.}\ \bibnamefont {Thielemann}},\ and\ \bibinfo
  {author} {\bibfnamefont {M.}~\bibnamefont {Wiescher}},\ }\href
  {https://doi.org/10.1088/0067-0049/189/1/240} {\bibfield  {journal} {\bibinfo
   {journal} {The Astrophysical Journal Supplement Series}\ }\textbf {\bibinfo
  {volume} {189}},\ \bibinfo {pages} {240} (\bibinfo {year}
  {2010})}\BibitemShut {NoStop}%
\bibitem [{\citenamefont {Cyburt}\ \emph {et~al.}(2016)\citenamefont {Cyburt},
  \citenamefont {Amthor}, \citenamefont {Heger}, \citenamefont {Johnson},
  \citenamefont {Keek}, \citenamefont {Meisel}, \citenamefont {Schatz},\ and\
  \citenamefont {Smith}}]{Cyburt2016}%
  \BibitemOpen
  \bibfield  {author} {\bibinfo {author} {\bibfnamefont {R.~H.}\ \bibnamefont
  {Cyburt}}, \bibinfo {author} {\bibfnamefont {A.~M.}\ \bibnamefont {Amthor}},
  \bibinfo {author} {\bibfnamefont {A.}~\bibnamefont {Heger}}, \bibinfo
  {author} {\bibfnamefont {E.}~\bibnamefont {Johnson}}, \bibinfo {author}
  {\bibfnamefont {L.}~\bibnamefont {Keek}}, \bibinfo {author} {\bibfnamefont
  {Z.}~\bibnamefont {Meisel}}, \bibinfo {author} {\bibfnamefont
  {H.}~\bibnamefont {Schatz}},\ and\ \bibinfo {author} {\bibfnamefont
  {K.}~\bibnamefont {Smith}},\ }\href
  {https://doi.org/10.3847/0004-637x/830/2/55} {\bibfield  {journal} {\bibinfo
  {journal} {The Astrophysical Journal}\ }\textbf {\bibinfo {volume} {830}},\
  \bibinfo {pages} {55} (\bibinfo {year} {2016})}\BibitemShut {NoStop}%
\bibitem [{\citenamefont {Meisel}\ \emph {et~al.}(2019)\citenamefont {Meisel},
  \citenamefont {Merz},\ and\ \citenamefont {Medvid}}]{Meisel_2019}%
  \BibitemOpen
  \bibfield  {author} {\bibinfo {author} {\bibfnamefont {Z.}~\bibnamefont
  {Meisel}}, \bibinfo {author} {\bibfnamefont {G.}~\bibnamefont {Merz}},\ and\
  \bibinfo {author} {\bibfnamefont {S.}~\bibnamefont {Medvid}},\ }\href
  {https://doi.org/10.3847/1538-4357/aafede} {\bibfield  {journal} {\bibinfo
  {journal} {The Astrophysical Journal}\ }\textbf {\bibinfo {volume} {872}},\
  \bibinfo {pages} {84} (\bibinfo {year} {2019})}\BibitemShut {NoStop}%
\bibitem [{\citenamefont {Parikh}\ \emph {et~al.}(2008)\citenamefont {Parikh},
  \citenamefont {Jos{\'{e}}}, \citenamefont {Moreno},\ and\ \citenamefont
  {Iliadis}}]{Parikh_2008}%
  \BibitemOpen
  \bibfield  {author} {\bibinfo {author} {\bibfnamefont {A.}~\bibnamefont
  {Parikh}}, \bibinfo {author} {\bibfnamefont {J.}~\bibnamefont {Jos{\'{e}}}},
  \bibinfo {author} {\bibfnamefont {F.}~\bibnamefont {Moreno}},\ and\ \bibinfo
  {author} {\bibfnamefont {C.}~\bibnamefont {Iliadis}},\ }\href
  {https://doi.org/10.1086/589879} {\bibfield  {journal} {\bibinfo  {journal}
  {The Astrophysical Journal Supplement Series}\ }\textbf {\bibinfo {volume}
  {178}},\ \bibinfo {pages} {110} (\bibinfo {year} {2008})}\BibitemShut
  {NoStop}%
\bibitem [{\citenamefont {Matic}\ \emph {et~al.}(2011)\citenamefont {Matic},
  \citenamefont {van~den Berg}, \citenamefont {Harakeh} \emph
  {et~al.}}]{Matic_2011}%
  \BibitemOpen
  \bibfield  {author} {\bibinfo {author} {\bibfnamefont {A.}~\bibnamefont
  {Matic}}, \bibinfo {author} {\bibfnamefont {A.~M.}\ \bibnamefont {van~den
  Berg}}, \bibinfo {author} {\bibnamefont {Harakeh}}, \emph {et~al.},\ }\href
  {https://doi.org/10.1103/PhysRevC.84.025801} {\bibfield  {journal} {\bibinfo
  {journal} {Phys. Rev. C}\ }\textbf {\bibinfo {volume} {84}},\ \bibinfo
  {pages} {025801} (\bibinfo {year} {2011})}\BibitemShut {NoStop}%
\bibitem [{\citenamefont {Meisel}\ and\ \citenamefont
  {Deibel}(2017)}]{Meisel_2017}%
  \BibitemOpen
  \bibfield  {author} {\bibinfo {author} {\bibfnamefont {Z.}~\bibnamefont
  {Meisel}}\ and\ \bibinfo {author} {\bibfnamefont {A.}~\bibnamefont
  {Deibel}},\ }\href {https://doi.org/10.3847/1538-4357/aa618d} {\bibfield
  {journal} {\bibinfo  {journal} {The Astrophysical Journal}\ }\textbf
  {\bibinfo {volume} {837}},\ \bibinfo {pages} {73} (\bibinfo {year}
  {2017})}\BibitemShut {NoStop}%
\bibitem [{\citenamefont {Lau}\ \emph {et~al.}(2018)\citenamefont {Lau},
  \citenamefont {Beard}, \citenamefont {Gupta} \emph {et~al.}}]{Lau18}%
  \BibitemOpen
  \bibfield  {author} {\bibinfo {author} {\bibfnamefont {R.}~\bibnamefont
  {Lau}}, \bibinfo {author} {\bibfnamefont {M.}~\bibnamefont {Beard}}, \bibinfo
  {author} {\bibfnamefont {S.~S.}\ \bibnamefont {Gupta}}, \emph {et~al.},\
  }\href {https://doi.org/10.3847/1538-4357/aabfe0} {\bibfield  {journal}
  {\bibinfo  {journal} {The Astrophysical Journal}\ }\textbf {\bibinfo {volume}
  {859}},\ \bibinfo {pages} {62} (\bibinfo {year} {2018})}\BibitemShut
  {NoStop}%
\bibitem [{\citenamefont {Morrissey}\ \emph {et~al.}(2003)\citenamefont
  {Morrissey}, \citenamefont {Sherrill}, \citenamefont {Steiner}, \citenamefont
  {Stolz},\ and\ \citenamefont {Wiedenhoever}}]{Morri2003}%
  \BibitemOpen
  \bibfield  {author} {\bibinfo {author} {\bibfnamefont {D.}~\bibnamefont
  {Morrissey}}, \bibinfo {author} {\bibfnamefont {B.}~\bibnamefont {Sherrill}},
  \bibinfo {author} {\bibfnamefont {M.}~\bibnamefont {Steiner}}, \bibinfo
  {author} {\bibfnamefont {A.}~\bibnamefont {Stolz}},\ and\ \bibinfo {author}
  {\bibfnamefont {I.}~\bibnamefont {Wiedenhoever}},\ }\href
  {https://doi.org/https://doi.org/10.1016/S0168-583X(02)01895-5} {\bibfield
  {journal} {\bibinfo  {journal} {Nuclear Instruments and Methods in Physics
  Research Section B: Beam Interactions with Materials and Atoms}\ }\textbf
  {\bibinfo {volume} {204}},\ \bibinfo {pages} {90 } (\bibinfo {year}
  {2003})}\BibitemShut {NoStop}%
\bibitem [{\citenamefont {Bradt}\ \emph {et~al.}(2017)\citenamefont {Bradt},
  \citenamefont {Bazin}, \citenamefont {Abu-Nimeh} \emph {et~al.}}]{Bradt2017}%
  \BibitemOpen
  \bibfield  {author} {\bibinfo {author} {\bibfnamefont {J.}~\bibnamefont
  {Bradt}}, \bibinfo {author} {\bibfnamefont {D.}~\bibnamefont {Bazin}},
  \bibinfo {author} {\bibfnamefont {F.}~\bibnamefont {Abu-Nimeh}}, \emph
  {et~al.},\ }\href
  {https://doi.org/https://doi.org/10.1016/j.nima.2017.09.013} {\bibfield
  {journal} {\bibinfo  {journal} {Nuclear Instruments and Methods in Physics
  Research Section A: Accelerators, Spectrometers, Detectors and Associated
  Equipment}\ }\textbf {\bibinfo {volume} {875}},\ \bibinfo {pages} {65 }
  (\bibinfo {year} {2017})}\BibitemShut {NoStop}%
\bibitem [{\citenamefont {Tarasov}\ and\ \citenamefont
  {Bazin}(2008)}]{Tarasov2008}%
  \BibitemOpen
  \bibfield  {author} {\bibinfo {author} {\bibfnamefont {O.}~\bibnamefont
  {Tarasov}}\ and\ \bibinfo {author} {\bibfnamefont {D.}~\bibnamefont
  {Bazin}},\ }\bibfield  {title} {\bibinfo {title} {Lise++: Radioactive beam
  production with in-flight separators},\ }\href
  {https://doi.org/https://doi.org/10.1016/j.nimb.2008.05.110} {\bibfield
  {journal} {\bibinfo  {journal} {Nuclear Instruments and Methods in Physics
  Research Section B: Beam Interactions with Materials and Atoms}\ }\textbf
  {\bibinfo {volume} {266}},\ \bibinfo {pages} {4657 } (\bibinfo {year}
  {2008})},\ \bibinfo {note} {proceedings of the XVth International Conference
  on Electromagnetic Isotope Separators and Techniques Related to their
  Applications}\BibitemShut {NoStop}%
\bibitem [{\citenamefont {Ayyad}\ \emph
  {et~al.}(2018{\natexlab{a}})\citenamefont {Ayyad}, \citenamefont {Abgrall},
  \citenamefont {Ahn} \emph {et~al.}}]{Ayyad2018}%
  \BibitemOpen
  \bibfield  {author} {\bibinfo {author} {\bibfnamefont {Y.}~\bibnamefont
  {Ayyad}}, \bibinfo {author} {\bibfnamefont {N.}~\bibnamefont {Abgrall}},
  \bibinfo {author} {\bibfnamefont {T.}~\bibnamefont {Ahn}}, \emph {et~al.},\
  }\bibfield  {journal} {\bibinfo  {journal} {Nuclear Instruments and Methods
  in Physics Research Section A: Accelerators, Spectrometers, Detectors and
  Associated Equipment}\ }\href
  {https://doi.org/https://doi.org/10.1016/j.nima.2018.10.019}
  {https://doi.org/10.1016/j.nima.2018.10.019} (\bibinfo {year}
  {2018}{\natexlab{a}})\BibitemShut {NoStop}%
\bibitem [{\citenamefont {Ayyad}\ \emph
  {et~al.}(2018{\natexlab{b}})\citenamefont {Ayyad}, \citenamefont {Mittig},
  \citenamefont {Bazin}, \citenamefont {Beceiro-Novo},\ and\ \citenamefont
  {Cortesi}}]{Ayyad_2018}%
  \BibitemOpen
  \bibfield  {author} {\bibinfo {author} {\bibfnamefont {Y.}~\bibnamefont
  {Ayyad}}, \bibinfo {author} {\bibfnamefont {W.}~\bibnamefont {Mittig}},
  \bibinfo {author} {\bibfnamefont {D.}~\bibnamefont {Bazin}}, \bibinfo
  {author} {\bibfnamefont {S.}~\bibnamefont {Beceiro-Novo}},\ and\ \bibinfo
  {author} {\bibfnamefont {M.}~\bibnamefont {Cortesi}},\ }\href
  {https://doi.org/https://doi.org/10.1016/j.nima.2017.10.090} {\bibfield
  {journal} {\bibinfo  {journal} {Nuclear Instruments and Methods in Physics
  Research Section A: Accelerators, Spectrometers, Detectors and Associated
  Equipment}\ }\textbf {\bibinfo {volume} {880}},\ \bibinfo {pages} {166 }
  (\bibinfo {year} {2018}{\natexlab{b}})}\BibitemShut {NoStop}%
\bibitem [{\citenamefont {Agostinelli}\ \emph {et~al.}(2003)\citenamefont
  {Agostinelli} \emph {et~al.}}]{Geant4}%
  \BibitemOpen
  \bibfield  {author} {\bibinfo {author} {\bibfnamefont {S.}~\bibnamefont
  {Agostinelli}} \emph {et~al.} (\bibinfo {collaboration} {GEANT4}),\
  }\bibfield  {title} {\bibinfo {title} {{GEANT4: A Simulation toolkit}},\
  }\href {https://doi.org/10.1016/S0168-9002(03)01368-8} {\bibfield  {journal}
  {\bibinfo  {journal} {Nucl. Instrum. Meth.}\ }\textbf {\bibinfo {volume}
  {A506}},\ \bibinfo {pages} {250} (\bibinfo {year} {2003})}\BibitemShut
  {NoStop}%
%%CITATION = NUIMA,A506,250;%%
\bibitem [{\citenamefont {Allison}\ \emph {et~al.}(2016)\citenamefont {Allison}
  \emph {et~al.}}]{geant41}%
  \BibitemOpen
  \bibfield  {author} {\bibinfo {author} {\bibfnamefont {J.}~\bibnamefont
  {Allison}} \emph {et~al.},\ }\bibfield  {title} {\bibinfo {title} {Recent
  developments in geant4},\ }\href
  {https://doi.org/https://doi.org/10.1016/j.nima.2016.06.125} {\bibfield
  {journal} {\bibinfo  {journal} {Nuclear Instruments and Methods in Physics
  Research Section A: Accelerators, Spectrometers, Detectors and Associated
  Equipment}\ }\textbf {\bibinfo {volume} {835}},\ \bibinfo {pages} {186 }
  (\bibinfo {year} {2016})}\BibitemShut {NoStop}%
\bibitem [{\citenamefont {Rauscher}(2010)}]{Raus10}%
  \BibitemOpen
  \bibfield  {author} {\bibinfo {author} {\bibfnamefont {T.}~\bibnamefont
  {Rauscher}},\ }\href {https://doi.org/10.1103/PhysRevC.81.045807} {\bibfield
  {journal} {\bibinfo  {journal} {\prc}\ }\textbf {\bibinfo {volume} {81}},\
  \bibinfo {pages} {045807} (\bibinfo {year} {2010})}\BibitemShut {NoStop}%
\bibitem [{\citenamefont {Galloway}\ \emph {et~al.}(2017)\citenamefont
  {Galloway}, \citenamefont {Goodwin},\ and\ \citenamefont {Keek}}]{Gall17}%
  \BibitemOpen
  \bibfield  {author} {\bibinfo {author} {\bibfnamefont {D.~K.}\ \bibnamefont
  {Galloway}}, \bibinfo {author} {\bibfnamefont {A.~J.}\ \bibnamefont
  {Goodwin}},\ and\ \bibinfo {author} {\bibfnamefont {L.}~\bibnamefont
  {Keek}},\ }\href@noop {} {\bibfield  {journal} {\bibinfo  {journal} {Publ.
  Astron. Soc. Aust.}\ }\textbf {\bibinfo {volume} {34}},\ \bibinfo {pages}
  {19} (\bibinfo {year} {2017})}\BibitemShut {NoStop}%
\bibitem [{\citenamefont {Wolf}\ \emph {et~al.}(2019)\citenamefont {Wolf},
  \citenamefont {Langer}, \citenamefont {Montes} \emph {et~al.}}]{Wolf19}%
  \BibitemOpen
  \bibfield  {author} {\bibinfo {author} {\bibfnamefont {C.}~\bibnamefont
  {Wolf}}, \bibinfo {author} {\bibfnamefont {C.}~\bibnamefont {Langer}},
  \bibinfo {author} {\bibfnamefont {F.}~\bibnamefont {Montes}}, \emph
  {et~al.},\ }\href {https://doi.org/10.1103/PhysRevLett.122.232701} {\bibfield
   {journal} {\bibinfo  {journal} {Phys. Rev. Lett.}\ }\textbf {\bibinfo
  {volume} {122}},\ \bibinfo {pages} {232701} (\bibinfo {year}
  {2019})}\BibitemShut {NoStop}%
\end{thebibliography}%

\end{document}